\documentclass[twocolumn,eqsecnum,showpacs,pra]{revtex4}
\usepackage{graphicx}

\begin{document}

\title{Atom-laser coherence and its control via feedback}

\author{L. K. Thomsen}
\affiliation{Centre for Quantum Dynamics, School of Science,
Griffith University, Brisbane, Queensland 4111, Australia}

\author{H. M. Wiseman}
\affiliation{Centre for Quantum Dynamics, School of Science,
Griffith University, Brisbane, Queensland 4111, Australia}

\date{\today}

\begin{abstract}

We present a quantum-mechanical treatment of the coherence
properties of a single-mode atom laser. Specifically, we focus on
the quantum phase noise of the atomic field as expressed by the
first-order coherence function, for which we derive analytical
expressions in various regimes. The decay of this function is
characterized by the coherence time, or its reciprocal, the
linewidth. A crucial contributor to the linewidth is the
collisional interaction of the atoms. We find four distinct
regimes for the linewidth with increasing interaction strength.
These range from the standard laser linewidth, through quadratic
and linear regimes, to another constant regime due to quantum
revivals of the coherence function. The laser output is only
coherent (Bose degenerate) up to the linear regime. However, we
show that application of a quantum nondemolition measurement and
feedback scheme will increase, by many orders of magnitude, the
range of interaction strengths for which it remains coherent.

\end{abstract}


\maketitle

\newcommand{\beq}{\begin{equation}}
\newcommand{\eeq}{\end{equation}}
\newcommand{\bqa}{\begin{eqnarray}}
\newcommand{\eqa}{\end{eqnarray}}
\newcommand{\ba}{\begin{array}}
\newcommand{\ea}{\end{array}}
\newcommand{\nn}{\nonumber}
\newcommand{\nl}[1]{\nn \\ && {#1}\,}
\newcommand{\erf}[1]{Eq.~(\ref{#1})}
\newcommand{\dg}{^\dagger}
\newcommand{\rt}[1]{\sqrt{#1}\,}
\newcommand{\smallfrac}[2]{\mbox{$\frac{#1}{#2}$}}
\newcommand{\half}{\smallfrac{1}{2}}
\newcommand{\bra}[1]{\langle{#1}|}
\newcommand{\ket}[1]{|{#1}\rangle}
\newcommand{\ip}[2]{\langle{#1}|{#2}\rangle}
\newcommand{\sch}{Schr\"odinger }
\newcommand{\schs}{Schr\"odinger's }
\newcommand{\hei}{Heisenberg }
\newcommand{\heis}{Heisenberg's }
\newcommand{\bl}{{\bigl(}}
\newcommand{\br}{{\bigr)}}
\newcommand{\ito}{It\^o }
\newcommand{\str}{Stratonovich }
\newcommand{\dpar}{\partial}
\newcommand{\dbd}[1]{\frac{\partial}{\partial {#1}}}
\newcommand{\deriv}[1]{\frac{d}{d {#1}}}
\newcommand{\sq}[1]{\left[ {#1} \right]}
\newcommand{\cu}[1]{\left\{ {#1} \right\}}
\newcommand{\ro}[1]{\left( {#1} \right)}
\newcommand{\an}[1]{\left\langle{#1}\right\rangle}
\newcommand{\implies}{\Longrightarrow}

\section{Introduction}

The invention of the laser in the late 1950s \cite{SchTow58}
created the field of quantum optics and continues to lead to an
enormous range of scientific and technological applications. It is
expected that the realization of ``atom lasers'' will similarly
revolutionize the field of atom optics \cite{alaserpot}. Atom
optics is the study of atoms where their wavelike nature becomes
important, suggesting an analogy with photons \cite{AdaSigMly94}.
An atom laser is therefore defined as a device that produces a
continuous beam of intense, highly directional, and
\textit{coherent} matter waves \cite{Wis97}, in analogy with the
light produced by an optical laser \cite{fn-continuous}. The ideal
atom laser beam is a single frequency (i.e., monochromatic) de
Broglie wave with well-defined intensity and phase.

The first experimental achievements of the Bose-Einstein
condensation (BEC) of gaseous atoms \cite{BECexpt} was followed
immediately by several independent ideas for creating an atom
laser \cite{proposals}. Since then there have been experimental
advances in the coherent release of pulses \cite{pulses} and
quasicontinuous beams \cite{beams} of atoms from BECs, as well as
further theoretical proposals \cite{props2}. In the experimental
configurations to date, the laser mode is the ground state of a
trapped BEC, which is pumped by evaporative cooling of uncondensed
atoms, and the out-coupling (separated in time from the pumping)
is achieved by either Raman or radio-frequency (rf) transitions to
an untrapped state. Although these experimental accomplishments do
not include simultaneous pumping and output coupling, they do
represent major steps towards achieving an operating atom laser.

Recent experimental \cite{KohHanEss01,Detetal01,LeCoqetal01} and
theoretical
\cite{Wis97,Doddetal97,ZobMey98,GolZobMey98,JacNarColWal99,HopMoyColSav00,RobSavOst01}
studies have focused on the fundamental coherence properties of
BECs and atom lasers. Atoms (unlike photons) interact with each
other, producing strong nonlinearities that affect the coherence
of the trapped condensate and thus also the out-coupled laser
field. For a single-mode condensate the dominant effect of atomic
collisions is to turn number fluctuations into fluctuations in the
energy and hence fluctuations in the frequency, thus causing
increased phase uncertainty. Collisional interactions therefore
lead to a significant decrease in the atom-laser coherence time,
and a corresponding increase in the linewidth, especially in the
case of BECs formed by evaporative cooling. However, we have
previously shown that a continuous, quantum nondemolition (QND)
feedback scheme can effectively cancel the linewidth broadening
due to such collisions \cite{WisTho01}.

To study the coherence properties of an atom laser one can either
focus on classical or quantum noise in the atomic field. Sources
of classical noise may be technical, such as fluctuations in the
trapping potential, finite temperatures, and specific trap
geometries \cite{Detetal01}, or dynamical, such as three-body
recombination \cite{RobSavOst01}. The study of these effects is
usually based on mean-field laser models described by
Gross-Pitaevskii (GP)-type equations \cite{LanLifPit80}. Quantum
noise is an intrinsic part of the atomic system as a consequence
of the uncertainty relations and is the limiting contribution to
coherence. The study of this requires a fully quantum-mechanical
approach, of which the most common is based on the quantum optical
master equation \cite{WalMil94}. The complexities of either
approach, which individually require approximations to facilitate
theoretical analysis, indicate that simultaneous analysis would
not be easy \cite{fn-hope}.

In this paper we present a fully quantum mechanical treatment of
the coherence of a single-mode atom laser and its control via the
QND feedback scheme proposed in Ref.~\cite{WisTho01}.
Specifically, we study the properties of the first-order coherence
function (i.e. phase coherence), which allows us to derive the
coherence time (and hence laser linewidth) as well as the power
spectrum of the laser output. Section II summarizes a set of
requirements for the coherence of an atom laser first detailed in
Ref.~\cite{Wis97}. Section III presents our mathematical model for
the atom laser and shows the resulting linewidth for increasing
atomic interaction strength. Numerical methods are presented in
Sec. III B, while the analytical results are discussed in Sec. III
C. When the collisional nonlinearity is very strong, the coherence
function undergoes a quantum collapse and revival sequence
(detailed in III D). This leads to an interesting regime in the
power spectrum (detailed in III E), which has not been considered
before. Section IV details the effects of feedback based on a
physically reasonable QND measurement of the condensate number, to
include all regimes of the nonlinearity. Section V concludes.

\section{Requirements for coherence}

The coherence of an atom-laser beam can be defined analogously to
that of an optical laser beam \cite{Wis97}. The fundamental
assumption is that the laser output is well approximated by a
highly directional classical wave of fixed intensity and phase,
which is also ideally restricted to a single transverse mode. The
output should also be a stationary process, i.e., its statistics
should be independent of time. To be coherent, the laser output
should then additionally have $(1)$ a relatively small spread of
longitudinal spatial frequencies (i.e. be monochromatic); $(2)$ a
relatively stable intensity (i.e. be approximately second-order
coherent); and $(3)$ a relatively stable phase (i.e. have a
relatively slow decay of first-order coherence).

The first condition, monochromaticity, follows from the
requirement that the laser output approximates a classical wave.
It can, therefore, be expressed in terms of the characteristic
coherence length of the wave, $l_{\rm coh}=(\delta k)^{-1}$, i.e.
the reciprocal of the spatial frequency spread. Condition (I)
becomes \beq l_{\rm coh}\gg\bar{\lambda}=2\pi/\bar{k}, \eeq i.e.,
that the coherence length be much greater than the mean atomic
wavelength \cite{Wis97}. In terms of the spectral width or
linewidth, $\ell\equiv\delta\omega$, the monochromaticity
requirement simply becomes $\ell\ll\bar{\omega}$, where the mean
frequency $\bar{\omega}$ is defined by the kinetic energy of the
atoms, i.e., $\bar{\omega}=\hbar\bar{k}^{2}/2m$. Condition (I)
also guarantees that the dispersion of an atomic beam will be
negligible over the coherence length \cite{Wis97}.

To explain the second and third conditions we require a many-body
description of the output beam; see Ref.~\cite{Wis97} for an
in-depth discussion. Basically, the output field of a laser can be
represented by the localized field annihilation operator $b(t)$,
which approximately satisfies the $\delta$-function commutation
relation
\beq [b(t),b\dg(t')]=\delta(t-t'),
\eeq
at a given point in the output. Thus, $I(t)=b\dg(t)b(t)$ can be
interpreted as the approximate atom-flux operator. The fundamental
assumption of a laser is that the output should be well
approximated by a classical wave of fixed intensity and phase.
This is therefore represented mathematically by
\beq
b(t)\approx\beta(t)\equiv\beta e^{-i\bar{\omega}t},
\label{fieldop}
\eeq
where $\beta$ is a complex number and the trivial time dependence
emphasizes that the laser output should be stationary. The atomic
field is not exactly a classical wave because there are
fluctuations in the field amplitude due to classical and quantum
sources of noise. These will need to be small or somehow cancelled
to maximize coherence.

It can be argued that there are no mean fields in both atom and
quantum optics \cite{Mol97}, and as such the description of a
laser as a coherent state is a ``convenient fiction''. Despite
this, mean-field theories (e.g., based on GP equations) are very
successful in describing the properties of lasers, as is the use
of initial coherent states for solving and visualizing master
equations (as done in this paper). A vanishing mean field means
that \beq \an{b(t)}=0. \label{0mean}\eeq In the case of optical
lasers the output might, in principle, be in a state with
well-defined amplitude and phase. But since we do not know the
absolute phase, an average over all possible phases gives
\erf{0mean}. For atom lasers, only bilinear combinations of the
atom field are observable and similarly no Hamiltonian is linear
in the atom field (atoms cannot be created out of nothing). Thus a
mean field amplitude is physically impossible and the absolute
phase is unobservable, again giving \erf{0mean}.

Since the mean field of a laser is zero, we cannot require
$\an{b(t)}=\beta$. However, for approximating $b(t)$ by
$\beta(t)$, we can require that the mean intensity be given by
\beq
\an{I(t)}=\an{b\dg(t)b(t)}=|\beta|^{2},
\label{avflux}
\eeq
and also that the fluctuations in intensity should be small in
some sense. This requirement is quantified using Glauber's normalized
second-order coherence function (for a stationary system)
\cite{Glau63}:
\beq
g^{(2)}(\tau)=\an{:I(t+\tau)I(t):}/\an{I(t)}^{2},
\label{g2}
\eeq
where the $\an{:\ :}$ denotes normal ordering.

For a field that is second-order coherent, i.e. $g^{(2)}(\tau)=1$,
there is no correlation between the arrival times of bosons at a
detector and their distribution is Poissonian. Specifically the
probability for detecting a boson in the interval
$(t+\tau,t+\tau+dt)$ given one detected at time $t$ is
$g^{(2)}(\tau)\an{I(t)}dt$. For the intensity fluctuations to be
small we therefore require that \cite{Wis97} \beq
|g^{(2)}(\tau)-1|\ll1, \label{two} \eeq i.e. the laser output
should be approximately second-order coherent: condition (II).

Assuming that condition (II) is met, the intensity of the laser
beam will be relatively stable and the only significant variation
in the output field will be due to phase fluctuations. A useful
measure of the phase fluctuations is the stationary first-order
coherence function \cite{Glau63}: \beq
G^{(1)}(\tau)=\an{b\dg(t+\tau)b(t)}, \eeq or its normalized form:
\beq g^{(1)}(\tau)=G^{(1)}(\tau)/\an{b\dg(t)b(t)}. \label{g1} \eeq
Unlike the field $b(t)$ itself, the bilinear combinations above
are measurable even for an atom field. $G^{(1)}(\tau)$ is simply
the mean intensity (\ref{avflux}) when $\tau=0$ and as $\tau$
increases, it decreases towards zero as the phase becomes
decorrelated from its initial value at $t$.

So, the phase of the field might be undefined because it varies in
time, i.e., the first-order coherence decays. Although we cannot
expect the laser to be approximately first-order coherent for all
time [i.e., $g^{(1)}(\tau)\approx 1$], we can require the decay of
$g^{(1)}(\tau)$ to be slow in some sense. The characteristic time
for this decay is simply the coherence time, which can be defined
as \cite{Wis97} \beq \tau_{\rm
coh}=\frac{1}{2}\int_{0}^{\infty}|g^{(1)}(\tau)|d\tau.
\label{cohtau} \eeq This is the time over which the phase of the
field is relatively constant.

But even if the phase is constant, it might also be undefined due
to a large intrinsic quantum uncertainty given by
$\Delta\phi\Delta n\geq1/2$ \cite{Hol84}. Since typically $\Delta
n\leq\bar{n}$, the quantum phase uncertainty will be large if the
mean number $\bar{n}$ is small. For the phase to be well defined
we therefore need the field to have a large intensity, i.e.,
$\bar{n}\gg1$, over the time that the phase of the field is
constant, i.e., for times $T\ll\tau_{\rm coh}$. The number of
bosons in the output field for a given duration $T$ is
$\bar{n}=\an{I(t)}T$. Thus, for a well-defined phase we require
$\an{I(t)}\tau_{\rm coh}\gg1$. In terms of the first-order
coherence function this translates to \cite{Wis97} \beq
\an{I(t)}\int|g^{(1)}(\tau)|d\tau=\int|G^{(1)}(\tau)|d\tau\gg1,
\label{three} \eeq which quantifies the requirement that the decay
of first-order coherence be relatively slow: condition (III).

This condition for coherence is equivalent to the requirement that
the output field be highly Bose-degenerate \cite{Wis97}, which is
rarely considered for optical lasers because it is so easily
satisfied. It requires that the output atom flux, i.e.,
$\an{I(t)}$, be much larger than the linewidth $\delta\omega$.
Since the atom-laser linewidth is the reciprocal of the coherence
time, i.e., \beq \ell\equiv\delta\omega=1/\tau_{\rm coh},
\label{linewidth} \eeq we find that condition (I) requires
$\ell\ll\bar{\omega}$ and condition (III) requires \beq
\ell\ll\an{I(t)}. \label{linethree} \eeq For single-mode optical
lasers typically $\an{I(t)}>\bar{\omega}$ so condition (III) is
always satisfied. On the other hand, the collisional interactions
in atom lasers causes significant linewidth broadening and so
satisfying (III) is not guaranteed. However, in Sec. IV we show
that this broadening can be effectively cancelled by a QND
measurement and feedback scheme.

The linewidth, or spectral width, of a field is usually defined as
the full width at half-maximum height (FWHM) of the output power
spectrum. In general, the power spectrum is given by the Fourier
transform of the first-order coherence function \cite{WalMil94}:
\beq
P(\omega)=\int_{-\infty}^{\infty}G^{(1)}(\tau)e^{-i\omega\tau}d\tau.
\label{PS} \eeq This is defined so that
$\int_{-\infty}^{\infty}P(\omega)d\omega=\an{I(t)}$ and therefore
can be interpreted as the steady-state flux per unit frequency
\cite{fn-pspec}.

If the first-order coherence function has the form
$G^{(1)}(\tau)\propto\exp(-\gamma\tau)$ then the laser output will
have a Lorentzian power spectrum. In this case the FWMH is exactly
equal to the linewidth as defined in \erf{linewidth}, i.e. the
reciprocal of the coherence time. On the other hand, if
$G^{(1)}(\tau)\propto\exp(-\gamma^{2}\tau^{2})$ then the laser has
a Gaussian power spectrum with a FWHM that is now only
approximately equal to the linewidth.

In any case, if the first-order coherence function has the form
$G^{(1)}(\tau)=|G^{(1)}(\tau)|\exp(i\bar{\omega}\tau)$, then
condition (III) for coherence can be restated in terms of the
maximum spectral intensity $P(\bar{\omega})$. From Eqs.~(\ref{PS})
and (\ref{cohtau}) and the above assumption (Appendix A shows that
this is a good approximation for the atom laser), we find \beq
P(\bar{\omega})=\int_{-\infty}^{\infty}|G^{(1)}(\tau)|d\tau=4\an{I(t)}\tau_{\rm
coh}. \label{PScentre} \eeq Thus, regardless of the resultant
shape of the output power spectrum, condition (III) becomes \beq
P(\bar{\omega})\gg1. \label{PSthree} \eeq Note that the central
frequency $\bar{\omega}$ will be shifted by any laser dynamics
which cause a rotation of the mean phase of the laser field.

The remaining sections of this paper present a study of the
quantum phase dynamics of the atom laser and thus the coherence
properties of its output. Specifically we study the first-order
coherence function as a measure of phase fluctuations, which (in the
absence of intensity fluctuations) will be the limiting factor to
the coherence time of an atom laser. This function, as indicated
in this section, is also intrinsically related to the laser output
characteristics of linewidth and power spectrum.

\section{Atom-laser linewidth}

\subsection{Atom-laser dynamics}

The atom-laser model consists of a source of atoms irreversibly
coupled to a laser mode, which is supported in a trap that allows
an output beam to form. A broadband reservoir acting both as a
pump and a sink is also coupled to the source modes. The laser
mode and source can be modelled by the ground and excited states
of a trapped boson field, representing the condensed and
uncondensed atoms, respectively \cite{Wis97}. Gain can be
achieved, for example, by evaporative cooling of the uncondensed
atoms. Out-coupling from the laser mode can be accomplished, for
example, by coherently driving condensed atoms into an untrapped
electronic state. This model can be simply described by a quantum
optical master equation for the laser mode alone
\cite{Wis97,Wis99}, obtained by adiabatically eliminating the
source modes and tracing over the continuum of output modes. The
system is thus characterized by a wavefunction $\Phi({\bf r})$ and
annihilation operator $a$ for the condensate mode.

Far above threshold, the laser mode has Poissonian number
statistics \cite{Lou73,SarScuLam74}. In the absence of thermal
or other excess noise, its dynamics are modelled by the completely
positive master equation \cite{Wis97,Wis99}
\beq
\dot{\rho} = \kappa\mu{\cal D}[a\dg]{\cal A}[a\dg]^{-1}\rho +
\kappa{\cal D}[a]\rho\equiv{\cal L}_{0}\rho,
\label{MEC0}
\eeq
where the superoperators ${\cal D}$ and ${\cal A}$ are defined as
usual for an arbitrary operator $r$:
\beq
{\cal D}[r]\rho \equiv r\rho r\dg - {\cal A}[r]\rho,~~
{\cal A}[r]\rho \equiv \half\{r\dg r,\rho\}.
\eeq
That the master equation is of the Lindblad form follows from
the identity
\beq
{\cal D}[a\dg]{\cal A}[a\dg]^{-1}
= \int_{0}^{\infty} dq{\cal D}[a\dg e^{-qaa\dg/2}].
\eeq

The first term of \erf{MEC0} represents linear output coupling at
rate $\kappa$ and the second term represents nonlinear (saturated)
pumping far above threshold, where $\mu\gg1$ is the stationary
mean boson number. It is the decreasing difference between the
gain and loss, as the laser is pumped above threshold, that gives
rise to the gain-narrowed laser linewidth \cite{WalMil94}. The
localized output mode operator $b$ of the preceding section is
then related to the laser mode via $b=\nu+\sqrt{\kappa}a$, where
$\nu$ represents vacuum fluctuations \cite{GarQnoise}. Note also
that we have chosen a reference potential energy for the system
such that there is no Hamiltonian $\propto a\dg a$ in the master
equation.

To include the effects of atom-atom interactions in the laser mode
we consider a simple $s$-wave scattering model for two-body
collisions, which is valid for low temperatures and densities
\cite{BalSavBECSS}. This is described by the Hamiltonian:
\beq
H_{\rm coll}=\hbar Ca\dg a\dg aa, ~~C=\frac{2\pi\hbar
a_{s}}{m}\int|\Phi({\bf r})|^{4}d^{3}{\bf r},
\label{Hcoll}
\eeq
where $\Phi({\bf r})$ is the condensate wavefunction and $a_{s}$
is the $s$-wave scattering length. The total master equation for
the laser mode including atomic interactions is then
\beq
\dot{\rho} = \kappa\mu{\cal D}[a\dg]{\cal A}[a\dg]^{-1}\rho +
\kappa{\cal D}[a]\rho-iC[a\dg a\dg aa,\rho]\equiv{\cal L}\rho,
\label{ME}
\eeq
where ${\cal L}$ is known as the Liouvillian for
the total system evolution.

As master equations, Eqs.~(\ref{MEC0}) and (\ref{ME}) are derived
using the Born-Markov approximation. More complicated mathematical
models for the atom laser dynamics may include a full multimode
description with non-Markovian pumping and/or damping (see, e.g.,
Refs.~\cite{JacNarColWal99,HopMoyColSav00}). However, only the
single-mode master equation description employed in this paper
allows a relatively straightforward analysis. Although a
single-mode scheme, e.g., \erf{ME}, ignores the many source modes
and the continuum of output modes, it is the simplest physically
reasonable model for an atom laser, in that it includes the
essential mechanisms of gain, loss and self-interactions.

There are also physical justifications for using Markovian theory.
Reference \cite{JacNarColWal99} states that the Markov
approximation is valid for output coupling rates satisfying
$\kappa^{-1}\gg T_m$, where $T_m$ is the output memory time, which
typically ranges from $10^{-2}$ to 1 ms. This condition is thus
easily satisfied for typical values of $\kappa^{-1}$, which range
from $10^{-2}$ to $10^{-1}$ s \cite{WisVac02}. Furthermore, as
discussed in Ref.~\cite{HopMoyColSav00}, the Born-Markov
approximation is only valid for either weak output coupling or
large atomic densities. BECs formed by evaporative cooling have
strong atom-atom interactions that correspond to large atomic
densities. We are therefore justified in making the Born-Markov
approximation in \erf{ME} [but not necessarily in \erf{MEC0}]. In
other words, we expect that strong nonlinear interactions, rather
than any non-Markovian dynamics, will dominate the linewidth.

\subsection{Numerical calculation of linewidth}

As stated in the Sec. II, the coherence time of a laser,
$\tau_{\rm coh}$, is roughly the time for the phase of the field
to become uncorrelated with its initial value. As shown by
\erf{cohtau}, it is determined by the stationary first-order
coherence function (\ref{g1}), in which the output operators $b$
can be replaced by the laser mode operators $a$, since
$b=\nu+\sqrt{\kappa}a$. For the evolution described by the master
equation (\ref{ME}), the coherence function becomes \beq
g^{(1)}(t)={{\rm Tr}[a\dg e^{{\cal L}t}a\rho_{\rm ss}]} /{{\rm
Tr}[\rho_{\rm ss}a\dg a]}, \label{cohfun} \eeq where $\rho_{\rm
ss}$ is the stationary solution to \erf{ME} given by
\cite{Lou73,SarScuLam74}: \beq \rho_{\rm
ss}=e^{-\mu}\sum_{n}\frac{\mu^{n}}{n!}\ket{n}\bra{n}
=\frac{1}{2\pi}\int_{0}^{2\pi}d\theta\ket{re^{i\theta}}\bra{re^{i\theta}},
\label{rhoss} \eeq where $r=\sqrt{\mu}$. Thus, the state of the
laser (\ref{rhoss}) can be thought of either as a mixture of
number states or equivalently a mixture of coherent states
(although see \cite{WisVac01} for a discussion of this).

It is a very good approximation (see Appendix A) to assume that if
$g^{(1)}(t)$ is not real then its complex nature is simply of the
form $g^{(1)}(t) = |g^{(1)}(t)|e^{i\bar{\omega}t}$ where
$\bar{\omega}$ is the central frequency of the laser output.
That is, we assume $g^{(1)}(t)$ is complex due
to an effective detuning ${\cal L}_{\bar{\omega}}=-i\bar{\omega}[a\dg
a,\rho]$ in the evolution. This type of evolution causes a
rotation of the mean phase proportional to $\bar{\omega}$, whereas
the decay of $|g^{(1)}(t)|$ indicates phase diffusion. Using
\erf{cohfun}, this allows the integral in \erf{cohtau} to be
evaluated to give
\beq
\tau_{\rm coh}\simeq-{{\rm Tr}[a\dg({\cal L}-i\bar{\omega})^{-1}a\rho_{\rm ss}]}
/{2{\rm Tr}[\rho_{\rm ss}a\dg a]}.
\label{numtau}
\eeq

Equation (\ref{numtau}) can be evaluated numerically, for example
using the {\sc Matlab} quantum optics toolbox \cite{Tan99}.
The first guess for $\bar{\omega}$ is found from the approximation
\beq
\bar{\omega}\simeq{\rm Im}\{{\rm Tr}[a\dg{\cal L}a\rho_{\rm ss}]\}\equiv\omega_{0},
\label{omega0}
\eeq
which is exact for ${\cal L}={\cal L}_{\bar{\omega}}$. Subsequent
corrections are found by an iterative procedure. Substituting
$\omega_{k}$ in \erf{numtau} gives $\tau_{k}$, which is then used
to update our guess for $\bar{\omega}$ via the expression
$\omega_{k+1}=\omega_{k}-{\rm Im}(1/2\tau_{k})$. Basically, this
scheme ensures that the calculated $\tau_{\rm coh}$ has a
vanishing imaginary component. This is justified in Appendix B,
where we also show that, if \erf{numtau} is valid, then only one
correction to $\omega_{0}$ is needed for an accurate determination
of ${\bar{\omega}}$.

\subsection{Analytical calculation of linewidth}

Analytically, it is easier to use the fact that \erf{cohfun} is
unchanged if $\rho_{\rm ss}$ is replaced by the initial coherent
state $\ket{re^{i\theta}}\bra{re^{i\theta}}$ for arbitrary
$\theta$ (say $\theta=0$). We therefore have \beq g^{(1)}(t)={\rm
Tr}[a\dg\rho(t)]/r,~~\rho(t)=e^{{\cal L}t}\ket{r}\bra{r}. \eeq
Using any suitable phase-space $(\alpha,\alpha^{*})$
representation, this expression is then equivalent to \beq
g^{(1)}(t)=\an{\alpha^{*}(t)}/\an{\alpha^{*}(0)}, \eeq where
$\an{\alpha^{*}(0)}=\an{\alpha(0)}=r$ and $\ket{\alpha}$ is a
coherent eigenstate of the laser field, i.e.,
$a\ket{\alpha}=\alpha\ket{\alpha}$. The state of the field at any
time can thus be described by the probability distribution for
$\alpha$, or equivalently (since $\alpha=\sqrt{n}e^{i\varphi}$)
the intensity, $n=|\alpha|^{2}$, and phase, $\varphi$,
distributions.

The fluctuations in intensity are relatively small for a laser
with $\mu\gg1$, i.e., $\delta n(t)\sim0$. Then, also assuming the
number statistics are unchanged by the evolution, we have
$n(t)\approx\bar{n}(t)=\bar{n}(0)$, which gives \beq g^{(1)}(t)=
\frac{\langle\sqrt{n(t)}e^{-i\varphi(t)}\rangle}{\langle\sqrt{n(0)}e^{-i\varphi(0)}\rangle}
\simeq \frac{\langle e^{-i\varphi(t)}\rangle}{\langle
e^{-i\varphi(0)}\rangle}. \eeq Now the phase distribution at time
$t$ due to the laser evolution is given by
$\varphi(t)=\varphi(0)+\phi(t)$, i.e., the phase distribution of
the initial coherent state plus the relative phase difference
$\phi(t)=\arg[\alpha(t)/\alpha(0)]$. Assuming this phase evolution
is independent of the initial phase uncertainty, we have \beq
g^{(1)}(t)\simeq\langle e^{-i\phi(t)}\rangle \approx
e^{-i\bar{\phi}(t)-\frac{1}{2}V_{\phi}(t)}, \label{g1phi} \eeq
where the second approximation assumes Gaussian statistics. The
coherence time, \erf{cohtau}, is thus found by evaluating the
integral: \beq \tau_{\rm coh} \approx
\frac{1}{2}\int_{0}^{\infty}e^{-\frac{1}{2}V_{\phi}(t)}dt.
\label{tauphi} \eeq

\begin{figure}
\includegraphics[width=0.25\textwidth]{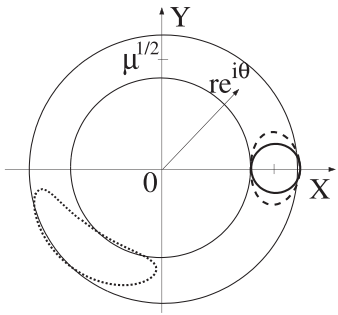}
\vspace{-0.3cm} \caption{\label{Qfnplot} Typical contour plots of
the $Q$ function for $\mu=15$ and
$C(\equiv\kappa\chi/4\mu)=\kappa(2\pi\mu)^{-1/2}$. Solid ring:
stationary laser state $\rho_{\rm ss}$, given by \erf{rhoss}.
Solid circle: initial coherent state
$\ket{\sqrt{\mu+1}}\bra{\sqrt{\mu+1}}$; dashed contour: phase
diffusion due to laser gain and loss at $t\sim0.8/\kappa$; dotted
contour: state due to total evolution including collisions, i.e.,
\erf{ME}, also at $t\sim0.8/\kappa$.}
\end{figure}

In the $Q$-function representation \cite{WalMil94}, a density
operator $\rho$ has a corresponding $Q$ function defined by \beq
Q(\alpha,\alpha^{*})=\bra{\alpha}\rho\ket{\alpha}/\pi, \eeq
normalized such that $\int Q(\alpha,\alpha^{*})d^{2}\alpha=1$. The
action of an operator on $\rho$ thus has the corresponding mirror
action of a differential operator acting on
$Q(\alpha,\alpha^{*})$. This is the most convenient representation
for our laser model because of the identity (see Appendix C) \beq
{\cal D}[a\dg]{\cal A}[a\dg]^{-1}\rho\to \sum_{k=1}^{\infty}\left(
-\dbd{n} \right)^{k}Q(n,\varphi), \label{gainQ} \eeq which, since
the higher order derivatives are negligible, can be truncated at
$k=2$. This representation also allows us to visualize the
dynamics produced by the master equation (\ref{ME}) as shown in
Fig.~\ref{Qfnplot}.

The master equation (\ref{ME}) thus turns into a Fokker-Planck
equation (FPE) for $Q(n,\varphi)$:
\bqa
\dbd{t}Q({\bf z},t)&=& \frac{1}{2}\sum_{j,k}\frac{\dpar^{2}}{\dpar z_{j}\dpar z_{k}}
[B_{jk}({\bf z})Q({\bf z},t)] \nl{-}\sum_{j}\dbd{z_{j}}[A_{j}({\bf z})Q({\bf z},t)],
\eqa
where ${\bf z}=(n,\varphi)$ and the drift vector $A$ and
diffusion matrix $B$ are given by
\beq
A=\left( \ba{c} \kappa(\mu+1-n) \\ (3-2n)C \ea \right),
~~
B=\left( \ba{cc} 2\kappa(\mu+n) & 2nC \\ 2nC & \kappa/2n \ea \right).
\eeq

To find equations of motion for the moments $\an{z_{j}}$ and
$\an{z_{j}z_{k}}$ the FPE needs to be converted to an
Ornstein-Uhlenbeck (OU) equation. For an OU process, the drift
vector is linear in the variables $(n,\varphi)$ and the diffusion
matrix constant. Our drift vector $A$ is already linear, but our
$B$ matrix is not constant. The simplest option is thus to replace
all the amplitudes in $B$ with their ($Q$-function) mean value,
i.e., $n\to\mu+1$. The equations of motion for the moments are
then $d\an{z_{j}}/dt=\an{A_{j}}$ and
$d\an{z_{j}z_{k}}/dt=\an{z_{j}A_{k}}+\an{z_{k}A_{j}}+(B_{jk}+B_{kj})/2$,
where $A$ and $B$ are now OU parameters.

We find that the number statistics are unchanged from that of the
initial coherent state, which for the $Q$-function representation
are $\bar{n}=\mu+1$ and $V_{n}=2\mu+1$. However, the phase-related
moments are altered. Using $\mu\gg1$, they are (for the relative
phase $\phi$)
\bqa
\bar{\phi}(t)&\simeq&-2\mu Ct, \label{phi} \\
V_{\phi}(t)&\simeq&\frac{8\mu C^{2}}{\kappa^{2}}\ro{e^{-\kappa t}+\kappa t-1}
+ \frac{\kappa}{2\mu}t, \label{vphi} \\
C_{n\phi}(t)&\simeq&\frac{2\mu C}{\kappa}\ro{e^{-\kappa t}-1}.
\label{covar} \eqa As expected, there is a mean phase shift
(\ref{phi}) due to the collisions, while the nonzero covariance
($C_{n\phi}=\an{n\phi} -\an{n}\an{\phi}$) given by \erf{covar}
explicitly shows the number-phase correlation produced by the
collisions. The phase variance (\ref{vphi}) contains two terms,
where the second corresponds to standard laser phase diffusion.
The first term thus indicates the increase in phase fluctuations
due to collisions.

The effect of collisions on the phase fluctuations can be clearly
seen in Fig.~\ref{Qfnplot}. This figure shows single contour plots
of the $Q$-function for the hypothetical coherent state
$\ket{\sqrt{\mu}}$ and snapshots at a later time due to the
evolution of the master equation (\ref{ME}). If we ignore
collisions the effect of the laser evolution is simply phase
diffusion. By including collisions we see two effects. First,
there is a rotation of the mean phase, due to \erf{phi}, and
second there is phase shearing. This is due to the nonzero
number-phase correlation, \erf{covar}, indicating that if the
inherent number fluctuations produce $n>\bar{n}$ the corresponding
phase $\phi$ will be less than $\bar{\phi}$, and vice versa. The
initial coherent state will approach the actual laser state
$\rho_{\rm ss}$, \erf{rhoss}, as $t\to\infty$.

Substituting \erf{vphi} into the expression for magnitude of the
first order coherence function (\ref{g1phi}) gives \beq
|g^{(1)}(t)|= e^{-\chi^{2}(e^{-\kappa t}+\kappa
t-1)/4\mu}e^{-\kappa t/4\mu}, \label{mg1phi} \eeq where we have
introduced $\chi=4\mu C/\kappa$ as a dimensionless parameter for
the atomic interaction strength. This expression does not have a
simple analytical solution. However, by inspection, there are two
limits that can be solved analytically. If $\chi\ll\sqrt{\mu}$ we
obtain \beq 2\tau_{\rm coh}\approx
\int_{0}^{\infty}e^{-\kappa(\chi^{2}+1)t/4\mu}dt
=4\mu/\kappa(1+\chi^{2}). \label{lorentz} \eeq For
$\chi\gg\sqrt{\mu}$, on the other hand, the first exponential in
\erf{mg1phi} is dominant and then expanding $e^{-\kappa t}$ to
second order we obtain \beq 2\tau_{\rm
coh}\approx\int_{0}^{\infty}e^{-\kappa^{2}\chi^{2}t^{2}/8\mu}dt
=\sqrt{2\pi\mu}/\kappa\chi. \label{gauss} \eeq

The resultant expression for the atom laser linewidth due to
collisions is thus
\beq
\ell
= \left\{ \begin{array}{lll}
    \kappa(1+ \chi^{2})/2\mu & ~{\rm for}~ & \chi \ll \sqrt{\mu} \\
    2\kappa \chi /\sqrt{2\pi\mu} & ~{\rm for}~ & \chi \gg \sqrt{\mu}
\end{array}  \right. .
\label{ellnofb} \eeq Clearly, for $\chi\ll1$, we obtain the
standard laser linewidth $\ell_{0}=\kappa/2\mu$
\cite{Lou73,SarScuLam74,Wis99}. These two expressions agree at
$\chi \simeq \sqrt{8\mu/\pi}$, and they are an excellent fit to
the numerical calculations of \erf{numtau}, except at the boundary
between the regimes. This is illustrated by the figure in our
previous paper \cite{WisTho01}, and also the extended version in
this paper, Fig.~\ref{linefb} appearing in Sec.~III D.

Equation (\ref{ellnofb}) represents the same physical dynamics as
found in similar studies by the authors of Refs.~\cite{ZobMey98}
and \cite{GarZoll98}. Zobay and Meystre \cite{ZobMey98} present a
three-mode atom-laser model, with the output mode adiabatically
eliminated. Ignoring collisions between pump and laser modes, they
obtain phase variances [Eqs. (21) and (22) of
Ref.~\cite{ZobMey98}], which are similar and identical to the
exponents of Eqs.~(\ref{lorentz}) and (\ref{gauss}) respectively.
See also the linewidth plotted in Fig.~3 of Ref.~\cite{ZobMey98}.
Gardiner and Zoller \cite{GarZoll98} studied a Bose-Einstein
condensate in dynamical equilibrium with thermal atoms. Our
first-order coherence function, \erf{mg1phi}, has the same
structure as the analogous expression, Eq. (184), derived in
Ref.~\cite{GarZoll98}. The two regimes of \erf{ellnofb} correspond
to the characteristic time constants of Eqs.~(187) and (186) of
Ref.~\cite{GarZoll98} respectively. The second of these
expressions, where the nonlinearity is dominant, is familiar as
the inverse collapse time of an initial coherent state in the
absence of pumping or damping \cite{WriWalGar96,ImaLewYou97}.

Since the output power spectrum is the Fourier transform of
$G^{(1)}(t)$ [see \erf{PS}], the shape of the spectrum is also
determined by the form of $V_\phi(t)$. For the two regimes of
\erf{ellnofb}, the laser output has Lorentzian and Gaussian power
spectra respectively, as was also found in Ref.~\cite{ZobMey98}.
These spectra are illustrated by Fig.~\ref{Pspecfb} in Sec.~IV B.
See Sec.~III C for a more in-depth discussion of the atom laser
power spectrum.

The standard laser linewidth (in the absence of collisions) is
simply given by $\kappa/2\mu$. For the preliminary atom laser
experiments of Refs.~\cite{pulses,beams}, the interaction strength
$C$ is always found to satisfy $C\gg\kappa/\mu$ and hence
$\chi\gg1$ \cite{WisVac02}. Atom lasers, therefore, have a
linewidth far above the standard limit. Furthermore, if
$\chi\agt\mu^{3/2}$ the linewidth will be larger than the mean
output flux $\kappa\mu$. In other words, the atomic collision
strength does not have to become very large before the laser
output does not satisfy condition (III) for coherence [i.e.,
\erf{linethree}]. It is thus of great interest to find methods for
reducing the linewidth due to atomic interactions. One method is
continuous QND measurement and feedback as shown in Sec.~IV.

\subsection{Revivals of the coherence function}

In the preceding section, the atom-laser linewidth was calculated
for atom-atom interactions ranging from weak ($\chi\ll1$) to
strong ($\chi\gg\sqrt{\mu}$). However, exact numerical
calculations based on \erf{numtau} indicated that there is an
upper bound to the linewidth (occurring for $\chi\agt\mu^{2}$)
that was not included in the previous analysis. It turns out that,
in addition to linewidth broadening, the collisional interactions
also lead to quantum revivals \cite{WriWalGar96} of the first
order coherence function. Although, note that in this very strong
collisional regime, the output atomic beam cannot be considered a
laser according to the definitions in Sec. II (since
$\ell\agt\kappa\mu$ for $\chi\agt\mu^{3/2}$).

To study the regime of revivals it is helpful to start by ignoring
all other laser dynamics apart from the collisions. In this case,
$\dot{\rho}=-iC[a\dg a\dg aa]\rho={\cal L}_{C}\rho$, and we can
analytically solve for the periodic structure of $g^{(1)}(t)$. For
an initial coherent state $\rho(0)=\ket{\alpha}\bra{\alpha}$,
\beq
g^{(1)}(t)={\rm Tr}[a\dg\rho(t)]/\alpha^{*},
~~\rho(t)=e^{{\cal L}_{C}t}\ket{\alpha}\bra{\alpha}.
\eeq

Using the number state representation, i.e. $\rho(t)=\sum
p_{mn}(t)\ket{n}\bra{m}$ and
$\ket{\alpha}=\exp(-|\alpha|^{2}/2)\sum
\alpha^{n}\ket{n}/\sqrt{n!}$, we find for the first-order
coherence function \beq g^{(1)}(t)=\exp\sq{-\mu(1-e^{2iCt})},
\label{g1coll} \eeq since $|\alpha|^{2}=\mu$ for the laser, and
its magnitude is \beq |g^{(1)}(t)|=\exp[-\mu(1-\cos2Ct)].
\label{mg1coll} \eeq The coherence function clearly has periodic
revivals when $t=m\pi/C$, $m$ is an integer.

Including the other laser dynamics, i.e. gain and loss, is not so
straightforward. Since these terms, unlike those in ${\cal
L}_{C}$, are not functions of the number operator we cannot easily
utilize the number state representation. The main effect, however,
is simply a decaying envelope applied to the revivals of
\erf{mg1coll}, such that the strength of $C$ compared to
$\kappa\mu$ will determine the number of significant revivals in
the coherence function. This will be shown below. The revivals of
the coherence function become significant when
$\chi\approx4\pi\mu^{2}$ or $C\approx\kappa\pi\mu$. This regime
was determined by calculating the exact linewidth based on
numerical solutions of \erf{numtau} and corresponds to the
interaction strength where the linewidth begins to approach a
maximum.

To determine the value of this maximum linewidth, we extend the
work of Milburn and Holmes \cite{MilHol86}, who model an
anharmonic oscillator coupled to a zero-temperature heat bath, via
two basic assumptions for including saturated gain. The master
equation modelled by Milburn and Holmes is (using our notation)
\beq \dot{\rho}_{\rm MH}=-iC[(a\dg a)^{2},\rho]+\kappa{\cal
D}[a]\rho, \eeq which gives a first-order coherence function of
the form \beq g^{(1)}_{\rm MH}(t) = e^{iCt}e^{-\kappa t/2} \exp
\sq{-\frac{\mu(1-i\kappa/2C)}{1+\kappa^{2}/4C^{2}}
\ro{1-e^{2iCt}e^{-\kappa t}} }. \eeq

To add saturated gain to this model we first assume that, far
above threshold, the contribution to phase diffusion is equal for
both gain and loss \cite{BarStePeg89} and so we replace $\kappa$
with $2\kappa$. Second, including gain will almost cancel the
overall exponential decay $e^{-\kappa t/2}$ of the coherence due
to loss, resulting in the smaller term $e^{-\kappa t/4\mu}$ (since
this will give the standard laser linewidth $\kappa/2\mu$). These
assumptions give the following results for $g^{(1)}(t)$ and
$|g^{(1)}(t)|$:
\begin{widetext}
\bqa
g^{(1)}(t)&\approx& e^{iCt}e^{-\kappa t/4\mu}
\exp \sq{ -\frac{\mu(1-i\kappa/C)}{1+\kappa^{2}/C^{2}}
\ro{1-e^{2iCt}e^{-2\kappa t}} }, \\
|g^{(1)}(t)|&\approx& e^{-\kappa t/4\mu} \exp \sq{
-\frac{\mu}{1+\kappa^{2}/C^{2}} \ro{1-e^{-2\kappa t}\cos2Ct
-\frac{\kappa}{C}e^{-2\kappa t}\sin2Ct.} } \eqa
\end{widetext}

We only expect this expression to be valid in the very strong
interaction regime ($C\gg\kappa\mu$). Here revivals of $|g^{(1)}(t)|$
are significant and $\kappa/C\ll1$ (since $\mu\gg1$). Also, as
shown by \erf{mg1coll}, revivals occur at $mt_{r}=m\pi/C$, so in
the strong regime the envelope of the coherence function is given
by (for finite $m$)
\bqa
|g^{(1)}(t)|_{\rm env}&=&|g^{(1)}(mt_{r})|
\simeq \exp \cu{-\ro{\frac{\kappa}{4\mu}+2\kappa\mu}mt_{r}} \nn \\
&\simeq& e^{-2t/t_{Q}},~~~t_{Q}=1/\kappa\mu.
\label{g1env}
\eqa
Here the time $t_{Q}$ can be interpreted as the quantum
dissipation time, i.e. how long the state would last if it was in
a superposition of coherent states. Nonunitary effects, such as
damping, cause a decay of the quantum coherence of these states at
a rate $\propto\kappa\mu$ \cite{WalMil85}. This is relevant
because the state produced halfway between revivals by nonlinear
interactions such as $\cal{L}_{C}$ is in fact a superposition of
coherent states \cite{DanMil89}.

The relationship between the quantum dissipation time and the
revival time can be used to give an indication of the number of
significant revivals in the coherence function for a given
interaction strength. That is, if $t_{r}\gg t_{Q}$ then no
revivals will be seen, but if $t_{r}\ll t_{Q}$ as for the above
equation, the number of significant revivals is of order
$t_{Q}/t_{r}$. Revivals begin to appear at $t_{r}\simeq t_{Q}$,
which is at $\chi\simeq 4\pi\mu^{2}$ or $C\simeq\kappa\pi\mu$ as
stated earlier.

We are now in a position to determine coherence time and linewidth
in the regime of revivals. From \erf{cohtau}, the coherence time
is simply half the area under the function $|g^{(1)}(t)|$. In the
very strong interaction regime ($C\gg\kappa\mu$), this area will
be made up of many individual peaks which decrease in height due
to the envelope given by \erf{g1env}. The first of these peaks
(which actually starts at $|g^{(1)}(0)|=1$) will be the same as
the coherence function for no revivals, and its area will be
$\sqrt{2\pi\mu}/\kappa\chi$ as given by \erf{gauss}. The
subsequent peaks will have areas twice this area multiplied by the
height of the envelope at that time. Thus we have for the total
area \bqa 2\tau_{\rm
coh}&\simeq&\frac{2\sqrt{2\pi\mu}}{\kappa\chi}
\cu{\int_{0}^{\infty}e^{-2t/t_{Q}}\sum_{m=0}^{\infty}\delta(t-mt_{r})dt-\frac{1}{2}}\nn \\
&=&\frac{2\sqrt{2\pi\mu}}{\kappa\chi}
\ro{\sum_{m=0}^{\infty}e^{-2mt_{r}/t_{Q}}-\frac{1}{2}}.
\eqa

\begin{figure}
\includegraphics[width=0.45\textwidth]{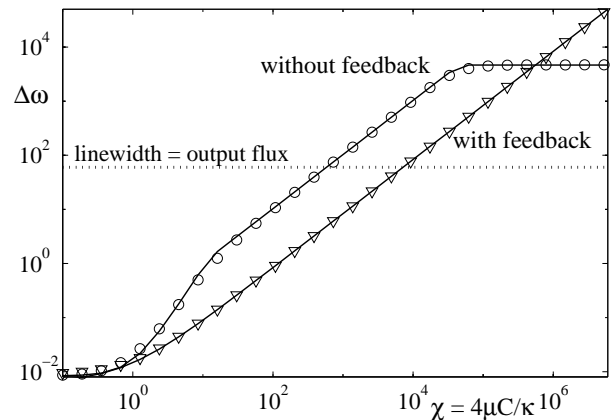}
\vspace{-0.4cm} \caption{\label{linefb} Atom laser linewidth in
units of $\kappa$ for $\eta=1$ and $\mu=60$, plotted with and
without feedback using both analytical (lines) and numerical
(points) methods. The dotted line corresponds to $\ell=\kappa\mu$,
i.e. interaction strengths with corresponding linewidths below
this line satisfy the coherence condition of Bose degeneracy
[condition (III)]. For the feedback results, see Section IV B.}
\end{figure}

This expression can be evaluated by using $t_{r}\ll t_{Q}$ and by
noting that for a geometric series, $\sum_{m}r^{m}=1/(1-r)$. The
analytical expression for the linewidth in the regime of revivals
is then \beq \ell_{\rm max}\simeq 4\kappa\sqrt{2\pi}\mu^{3/2}.
\label{ellmax} \eeq This is in exact agreement with the numerical
results obtained from \erf{cohfun}, which are plotted
Fig.~\ref{linefb}.

In this figure we have plotted the approximate analytical
expressions for the linewidth in the absence [Eqs.~(\ref{ellnofb})
and (\ref{ellmax})] and presence [\erf{ellfb}] of feedback (see
Sec.~IV B for details) as a function of the nonlinearity $\chi$
for $\mu=60$. We have also included numerical results (see Sec.
III B for details) as a comparative test for the analytical work,
which is valid for $\mu\gg1$. As can be seen, the agreement is
very good even for an occupation number of only 60
\cite{fn-compmem}, thus confirming the accuracy of our analytical
expressions for the linewidth. Without feedback we see four
distinct regimes. There is the standard laser linewidth for
$\chi\ll1$, a quadratic dependence on $\chi$ for
$1\ll\chi\ll\sqrt{\mu}$, and a linear regime for
$\sqrt{\mu}\ll\chi\ll\mu^{2}$. The latter two correspond to the
regimes of \erf{ellnofb}. Finally there is a constant regime given
by \erf{ellmax} for $\chi\gg\mu^{2}$, which is due to the
collapses and revivals of the coherence function.

Note that the approximation used in the numerical calculation of
\erf{numtau}, i.e., $g^{(1)}(t)=|g^{(1)}(t)|e^{i\bar{\omega}t}$,
is no longer necessarily valid in the regime of revivals, as
indicated by the multi-complex-exponential nature of
\erf{mg1coll}. Nevertheless, our numerical results are still
correct because we have taken the mean atom number $\mu$ to be an
integer. At revivals the approximation to $g^{(1)}(t)$ becomes
\beq
|g^{(1)}(mt_{r})|e^{i\omega_{0}mt_{r}}=|g^{(1)}|e^{i2m\pi\mu},
\eeq where we have only used the first guess for $\bar{\omega}$,
since the iterative procedure will be inaccurate in this regime
(see Appendix B). This expression clearly equals $|g^{(1)}|$ for
integer $\mu$. Thus, the same numerical simulation can be used for
all values of $C$ as long as $\mu$ is an integer

\subsection{Power spectrum}

As stated in Sec. II, condition (III) for coherence requires that
the integral of $|G^{(1)}(t)|$ be much greater than unity
[\erf{three}]. This was reinterpreted as requiring the linewidth
to be much less than the output flux (Bose degeneracy), or
equivalently requiring the maximum spectral intensity to be much
greater than unity $P(\bar{\omega})\gg1$. Now the linewidth is
only the FWHM of the power spectrum if it is Lorentzian. As
discussed after \erf{ellnofb}, this will only be the case in the
weak-interaction regime $\chi\ll\sqrt{\mu}$. As $\chi$ (or $C$) is
increased the power spectrum becomes Gaussian, and as $\chi$
enters the very strong interaction regime it will no longer have a
simple structure at all. At these strong values of the
nonlinearity we see quantum revivals of $g^{(1)}(t)$.

In terms of normalized first-order coherence function the power
spectrum becomes \beq
P(\omega)=\kappa\mu\int_{-\infty}^{\infty}g^{(1)}(t)e^{-i\omega
t}dt, \label{PSg1} \eeq where we have recognized that
$\an{I}=\an{b\dg b}= \kappa\an{a\dg a}=\kappa\mu$. From this
equation it is clear that as long as $g^{(1)}(t)$ has a simple
structure, i.e., no revivals, then the spectrum will have a simple
(Lorenztian or Gaussian) lineshape for a given interaction
strength $C$, with the intensity and width determined by how fast
$g^{(1)}(t)$ decays.

The maximum spectral intensity was defined in Sec. II by
\erf{PScentre}, which was based on the assumption that
$g^{(1)}(t)=|g^{(1)}(t)|e^{i\bar{\omega}t}$. Since we are in a
reference frame with zero mean frequency before including
collisions, $\bar{\omega}$ is the detuning frequency due to
collisions which causes the mean phase shift of an initial
coherent state (see Fig.~\ref{Qfnplot}). At this frequency, the
output power spectrum will, therefore, have a maximum value given
by \beq P(\bar{\omega})=4\kappa\mu\tau_{\rm coh}. \eeq As stated
above, in the regime of revivals this approximation will only be
accurate for the first guess for $\bar{\omega}$, i.e.,
$\omega_{0}=2\mu C$.

The linewidth in the strong atomic interaction regime
($C\agt\kappa\pi\mu$) was calculated in the preceding section to
be $4\kappa\sqrt{2\pi}\mu^{3/2}$, i.e. \erf{ellmax}. Thus, from
the above equation, we expect the maximum spectral intensity to
reach \beq P(\omega_{0})=4\kappa\mu/\ell_{\rm
max}=1/\sqrt{2\pi\mu}, \label{PSmax} \eeq as the atomic
interaction strength $C$ is increased far above $\kappa\pi\mu$.

\begin{figure}
\includegraphics[width=0.47\textwidth]{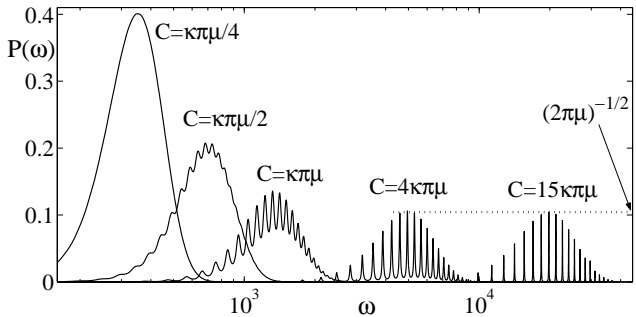}
\vspace{-0.2cm} \caption{\label{Pspec} Output power spectrum
$P(\omega)$ where $\omega$ has units of $\kappa$, plotted for
$\mu=15$ and various values of $C$ in the strong interaction
regime. The dotted line corresponds to the value
$4\kappa\mu/\ell_{\rm max}$, i.e., the maximum height of the power
spectrum in the regime of revivals. Note the log scale for
$\omega$.}
\end{figure}

Figure~\ref{Pspec} illustrates both the simple lineshape of the
power spectrum when there are no revivals and the complication of
the spectrum as $C$ is increased. Since all the plotted spectra
have $P(\bar{\omega})<1$, the atom-laser output clearly does not
satisfy condition (III) for coherence, \erf{PSthree}, in the
strong interaction regime. The first spectrum in Fig.~\ref{Pspec}
is in the range $\sqrt{\mu}<C<\kappa\pi\mu$, and thus although
revivals are not seen, the output is still above the cutoff for
Bose degeneracy.

The remaining spectra illustrate the increasing effect of quantum
revivals due to increasing interaction strength. The central peak
of the power spectrum, which can be defined regardless of revivals
as shown in \erf{PScentre}, clearly approaches the predicted
maximum of $1/\sqrt{2\pi\mu}$, i.e., \erf{PSmax}, for
$C\gg\kappa\pi\mu$. For examples of the spectrum in the weak
interaction regimes of \erf{ellnofb}, see Sec.~IV B.

\section{Reducing the linewidth via feedback}

Section III A showed that the atomic interactions do not directly
cause phase diffusion. Rather, they cause a shearing of the field
in phase space, with higher amplitude fields having higher energy
and hence rotating faster. The resultant linewidth broadening is a
known effect for optical lasers with a Kerr ($\chi^{(3)}$) medium
\cite{Watetal90}. The shearing of the field is manifest in the
finite value acquired by the covariance $C_{n\phi}(t)$ in
\erf{covar}. It means that information about the condensate number
is also information about the condensate phase. Hence, we can
expect that feedback based on atom number measurements could
enable the phase dynamics to be controlled, and the linewidth
reduced.

\subsection{QND feedback scheme}

\begin{figure}
\includegraphics[width=0.35\textwidth]{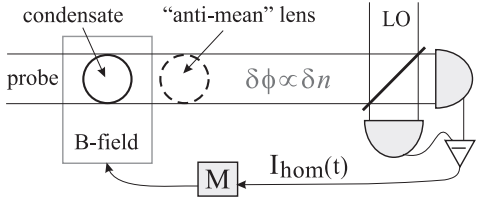}
\vspace{-0.2cm} \caption{\label{exptfb} Experimental schematic. A
far-detuned probe laser of amplitude $\varrho$ interacts with the
condensate (i.e., the laser cavity) causing a phase shift in the
probe proportional to the condensate number. The ``anti-mean''
lens subtracts the mean phase shift leaving the probe with a phase
shift proportional to the number fluctuations $\delta n$. The
photocurrent $I_{\rm hom}(t)$ from the homodyne $Y$ detection of
this field is then used to modulate an external $B$-field
uniformly applied to the condensate.}
\end{figure}

QND atom number measurements can be performed on the condensate
{\em in situ} via the homodyne detection of a far-detuned probe
field \cite{Andetal96,CorMil98,Lyeetal99,Daletal01}. This
dispersive interaction causes a phase shift of the probe
proportional to the number of atoms in the condensate. We consider
a far-detuned probe laser beam of frequency $\omega_p$ and
cross-sectional area $A$ that passes through the condensate
\cite{fn-antimean}. Figure \ref{exptfb} shows an experimental
schematic for our QND measurement and feedback scheme.

For a single atom the interaction with the optical probe field can
be approximated by the Hamiltonian
\beq
V=\hbar\frac{\Omega_p^{2}}{4\Delta} = \hbar
\left( \frac{\hbar\omega_p\gamma^{2}}{8A\Delta I_{\rm sat}}\right)p\dg p
\equiv \hbar\theta p\dg p ,
\label{deftheta}
\eeq
where $\Omega_{p}$, $\Delta$, $\gamma$ have their usual meaning
and $I_{\rm sat}=2\pi hc\gamma/\lambda^3$ \cite{Ash70}. Here $p$ is
the annihilation operator for the probe beam, normalized so that
$\hbar \omega_p p\dg p$ is the beam power.

The effective interaction Hamiltonian for the whole
condensate can thus be taken to be
\beq
H_{\rm int} = \hbar \theta (a\dg a-\mu) p\dg p,
\label{Hint}
\eeq
where $\theta$, defined in \erf{deftheta}, is the phase shift of the
probe field due to a single atom. Here we have also subtracted the
mean phase shift such that the probe optical laser is a measure of
the atom number fluctuations only (see Fig.~\ref{exptfb}).

The back action on the condensate due to this interaction can be
evaluated using the techniques of Sec.~III B of Ref.~\cite{Wis94}.
Assuming the input probe field is in a coherent state of amplitude
$\varrho$ and mean power $P$, the evolution of the atomic system
due the measurement is \beq \dot{\rho} =\varrho^{2}{\cal
D}[e^{-i\theta(a\dg a-\mu)}]\rho \simeq M {\cal D}[a\dg a]\rho,
\label{MEmeas} \eeq where the the measurement strength is given by
\beq M = \varrho^{2}\theta^{2}=P\theta^{2}/\hbar\omega_p. \eeq The
approximation in \erf{MEmeas} requires $\theta(a\dg a-\mu)\ll1$,
which for Poissonian number fluctuations is simply
$\sqrt{\mu}\theta \ll 1$.

The above result represents decoherence of the atom laser due to
photon number fluctuations in the probe field, resulting in
increased phase noise. In a recent theoretical study by Dalvit and
co-workers \cite{Daletal01}, it was shown that dispersive
measurements of BECs cause both phase diffusion [as in
\erf{MEmeas}] and atom losses. Nevertheless, they also show that
phase diffusion dominates the decoherence rate for large atom
numbers, i.e., for $\mu\gg1$, and so the depletion contribution
can generally be neglected \cite{fn-depletion}. Equation
(\ref{MEmeas}) is equivalent to the corresponding phase diffusion
term in their work, since it can be shown that both $M$ and the
phase diffusion rate given by Eq.~(14) of \cite{Daletal01} reduce
to $\sim\gamma^{2}\lambda^{5}I/hcA\Delta^{2}$.

The effect of the interaction (\ref{Hint}) on the output probe
field is to cause a phase shift proportional to the number
fluctuations, $a\dg a-\mu$. The output field operator is given
by \cite{Wis94}
\beq
p_{\rm out} = e^{-i\theta (a\dg a-\mu)}p_{\rm in} \simeq p_{\rm in}
- i\varrho\theta(a\dg a-\mu),
\eeq
where again the approximation requires $\sqrt{\mu}\theta \ll 1$.
Homodyne detection of the $Y$ quadrature of the output probe
field will thus be a measure of the condensate number fluctuations.
The homodyne photocurrent operator is given by \cite{Wis94}
\beq
I^{\rm Y}_{\rm out} = -ip_{\rm out}+ip_{\rm out}\dg
\simeq I^{\rm Y}_{\rm in} -2\sqrt{M}(a\dg a - \mu).
\eeq

In order to control the phase dynamics of the condensate, we wish
to use this homodyne current to modulate its energy. This can be
done, for example, by applying a uniform magnetic field or
far-detuned light field across the whole condensate. In the
ideal limit of instantaneous feedback, we model this
by the Hamiltonian
\beq
H_{\rm fb}(t) = -\hbar a\dg a FI_{\rm hom}(t)/\sqrt{M},
\eeq
where $F$ is the feedback strength and $I_{\rm hom}(t)$ is the
classical photocurrent corresponding to the operator
$I^{\rm Y}_{\rm out}$.

The total evolution of the system including feedback is obtained
by applying the Markovian theory of Ref.~\cite{Wis94}. The master
equation becomes
\bqa
\dot\rho &=& {\cal L}_{0}\rho -iC[a\dg a\dg aa,\rho] + M{\cal D}[a\dg a]\rho
\nl{+} iF[a\dg a\dg a a,\rho] + \frac{F^{2}}{\eta M}{\cal D}[a\dg a]\rho,
\label{MEfb}
\eqa
where we have allowed for a detection efficiency $\eta$
\cite{Wis94} and dropped terms corresponding to a frequency shift.

\begin{figure}
\includegraphics[width=0.35\textwidth]{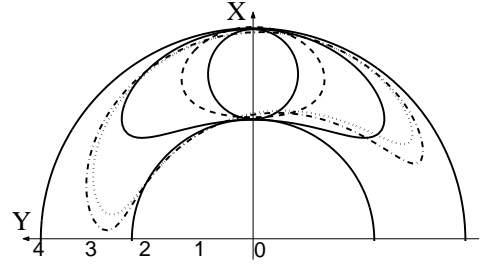}
\vspace{-0.2cm} \caption{\label{Qplotfb} Contour plots of the $Q$
function for $\mu=10$ and $C=\kappa(2\pi\mu)^{-1/2}$ (ignoring any
mean phase shifts). Black circle: initial coherent state
$\ket{\sqrt{\mu}}\bra{\sqrt{\mu}}$; black ring: stationary laser
state $\rho_{\rm ss}$, \erf{rhoss}. The other contours correspond
to the evolution at time $t\sim 1.5/\kappa$ due to successive
terms in the master equation. Dashed contour: phase diffusion due
to ${\cal L}_{0}$; dotted contour: including atomic collisions,
i.e., ${\cal L}_{0}$ and $C$; dot-dash contour: including QND
measurement back action, i.e., ${\cal L}_{0}$ and $ C$ and $M$,
and finally the solid contour corresponds to \erf{MEopt}. All
contours are for the optimal feedback regime $F=\sqrt{\eta}M=C$.}
\end{figure}

The terms in \erf{MEfb} describe, respectively, the standard laser
gain and loss (${\cal L}_{0}$), the collisional interactions
($C$), the measurement back action ($M$), the feedback phase
alteration ($F$), and the noise introduced by the feedback. As
before we can visualize the effect of the measurement and feedback
terms on the evolution of an arbitrary coherent state. This is
illustrated by the $Q$-function contours in Fig.~\ref{Qplotfb}. In
this figure we have ignored the mean phase shift due to collisions
to make the comparison clearer.

To completely remove that unwanted nonlinearity, the obvious choice
for the feedback strength is $F=C$. We also want to minimize the
phase diffusion introduced by both the measurement and feedback. A
weak measurement will give poor information about the atom number,
with a high noise-to-signal ratio, which will increase the noise
due to feedback. On the other hand, if the measurement is too
strong the measurement back action itself will dominate. This
leads us to guess the optimal regime for both measurement and
feedback to be $F=\sqrt{\eta}M=C$, which simply leaves
\beq
\dot\rho = \kappa\mu{\cal D}[a\dg]{\cal A}[a\dg]^{-1}\rho +
\kappa{\cal D}[a]\rho + \frac{2C}{\sqrt{\eta}}{\cal D}[a\dg a]\rho.
\label{MEopt}
\eeq

\subsection{Linewidth results}

Proceeding as before, we can find the exact effect of the general
feedback scheme on the atom-laser linewidth. Neither the
measurement nor the feedback affect the atom number statistics.
The change in the phase statistics are reflected by the new
Fokker-Planck equation for $Q(n,\phi)$. The altered terms in the
drift vector and diffusion matrix are $A_{2}=(3-2n)(C-F)$,
$B_{12}=B_{21}=2n(C-F)$, and $B_{22}=\kappa/2n+M+F^{2}/\eta M$.
After linearizing, we again find the phase-related moments: \bqa
\bar{\phi}(t)&\simeq&-2\mu (C-F)t, \label{phifb} \\
V_{\phi}(t)&\simeq&\frac{8\mu (C-F)^{2}}{\kappa^{2}}\ro{e^{-\kappa t}+\kappa t-1}
\nl{+} \ro{\frac{\kappa}{2\mu}+M+\frac{F^{2}}{\eta M}}t, \label{vphifb1} \\
C_{n\phi}(t)&\simeq&\frac{2\mu (C-F)}{\kappa}\ro{e^{-\kappa t}-1}, \label{covarfb}
\eqa
where again the approximations have used $\mu\gg1$.

These equations clearly show that all the unwanted phase
statistics are cancelled by choosing a feedback regime with $F=C$
and furthermore the minimum phase variance is when
$M=C/\sqrt{\eta}$. Specifically,
both the mean phase shift and the correlation between number and
phase fluctuations are removed, and the phase variance
is simply given by
\beq
V_{\phi}(t)= \ro{\frac{\kappa}{2\mu}+\frac{2C}{\sqrt{\eta}}}t.
\label{vphifb}
\eeq
This is exactly the phase variance from the master equation
(\ref{MEopt}).

In this case, \erf{tauphi} has a simple analytical solution: \beq
\ell=\tau_{\rm
coh}^{-1}=\frac{\kappa}{2\mu}\ro{1+\frac{\chi}{\sqrt{\eta}}},
\label{ellfb} \eeq where we have again used the dimensionless
atomic interaction strength $\chi=4\mu C/\kappa$. Note that unlike
\erf{ellnofb}, this linewidth is valid for all $\chi$. The
derivation of \erf{ellfb} above is based on preselecting the
feedback and measurement parameters (i.e., $F=\sqrt{\eta}M=C$). On
the other hand, Ref.~\cite{WisTho01} presented the analytical
solution for the linewidth independent of the choice of these
parameters and proceeded to find the minimum with respect to the
feedback strength. The result [Eq.~(24) of Ref.~\cite{WisTho01}]
only differs from \erf{ellfb} by the term $\kappa(-1/4\eta)/2\mu$.

From Fig.~\ref{linefb} (in Sec.~III C) and the above result,
\erf{ellfb}, it is evident that our QND feedback scheme offers a
linewidth much smaller than that without feedback for most values
of $\chi$. In fact for $\sqrt{\mu}\alt\chi\alt4\pi\mu^{2}$ the
reduction in linewidth due to feedback is a factor of
$\sqrt{8\mu/\pi}$. Most importantly, the laser output is Bose
degenerate [satisfies condition (III) for coherence], up to
$\chi\approx\mu^{2}$ with feedback, as opposed to
$\chi\approx\mu^{3/2}$ in the absence of feedback. Thus, the atom
laser with feedback remains coherent for much stronger atomic
nonlinearities than without feedback. It is interesting that this
corresponds to the ``conditionally coherent'' regime
$\mu^{3/2}<\chi<\mu^{2}$ as discussed in Ref.~\cite{WisVac02}.

Since the nonlinearity $C$ is effectively cancelled by this
feedback scheme, the output power spectrum, \erf{PSg1}, will never
have the complicated structure as shown in Sec.~III C. Also, since
\erf{vphifb} has a linear dependence on time (rather than higher
powers) the first-order coherence function decays exponentially
and as such will produce a Lorentzian output power spectrum [see
discussion after \erf{PS}].

\begin{figure}
\includegraphics[width=0.46\textwidth]{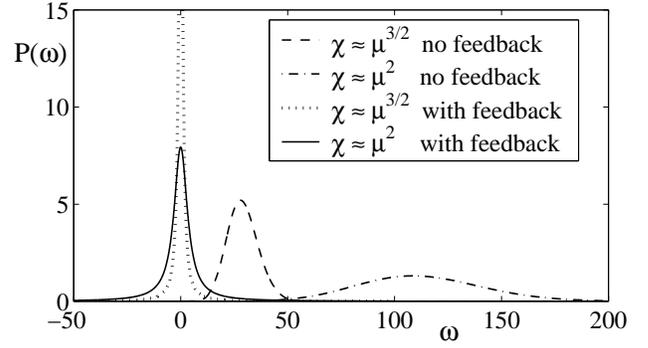}
\vspace{-0.4cm} \caption{\label{Pspecfb} Output power spectrum
$P(\omega)$ for $\mu=15$ vs $\omega$ in units of $\kappa$, plotted
with and without feedback for interaction strengths of
$\chi=\mu^{3/2}$ and $\chi=\mu^{2}$. The dotted spectrum
($\chi=\mu^{3/2}$ with feedback) has a maximum intensity of
$\sim31$.}
\end{figure}

This lineshape is illustrated in Fig.~\ref{Pspecfb}, which also
plots the corresponding spectra that would be produced by the
laser (for the same values of $\chi$) without feedback. These
latter spectra have a Gaussian lineshape as discussed after
\erf{ellnofb}. Also, as indicated by \erf{phifb}, the rotation of
the mean phase due to collision is cancelled by our feedback
scheme, i.e., $g^{(1)}(t)$ is no longer complex and hence
$\bar{\omega}=0$. The power spectrum for the atom laser including
feedback will thus be centered around zero frequency regardless of
the atomic interaction strength, which is confirmed in the figure.

The spectra in Fig.~\ref{Pspecfb} are plotted at the two cutoff
values for Bose degeneracy, which are $\chi\approx\mu^{3/2}$ for
no feedback and $\chi\approx\mu^{2}$ when feedback is included. If
the output is coherent, i.e., satisfies Bose degeneracy, then it
clearly has a much narrower and thus more intense spectrum than if
the output is not coherent. This figure thus illustrates the link
between the three definitions of linewidth [discussed after
\erf{PScentre}] and its relation to coherence. That is, a narrow
intense output spectral line corresponds to a long phase coherence
time, and more specifically, if the FWHM linewidth is much less
than the value of the output flux, then the laser output will
satisfy the conditions of coherence.

\subsection{Experimental realizability}

We will now briefly examine some issues of experimental
realizability. A question of interest is, how easy is it to obtain
a QND measurement of sufficient strength to optimize the feedback?
We have shown in Sec.~IV B that the optimum feedback scheme
requires $N=C/\sqrt{\eta}$. From Ref.~\cite{WisVac02} we know that
most current experiments work in the regime where the Thomas-Fermi
approximation can be made, allowing an analytical expression to be
found for $\chi$ [Eq.~(6.25) in \cite{WisVac02}] and $C$. Typical
values are $\chi\sim10^{3}$, $\mu\sim10^6$ and
$C\sim10^{-2}$s$^{-1}$. To determine $\theta$ and hence $M$ we use
typical $^{87}$Rb imaging parameters \cite{Lyeetal99,Daletal01},
which include $\lambda=780$nm, $A\sim10^{-11}$m$^{2}$,
$\gamma\sim5$Mhz, $\Delta\sim2$GHz, and $I_{\rm
sat}\sim10$W$/$m$^{2}$. For these values,
$\theta\sim3.3\times10^{-6}$, and thus $M\sim
4.2\times10^{-4}\times I$. To obtain a measurement strength of the
order of $C$, we therefore require a probe laser intensity of only
$\sim30$W/m$^{2}$, which is quite reasonable.

A related question is, how much of a problem is atom loss due to
spontaneous emission by atoms excited by the detuned probe beam?
The rate of this loss (ignoring reabsorption) is $\gamma\times
(\textrm{excited population})$. We would like the ratio of this
loss rate to the output loss rate $\kappa\mu$ to be small. In the
optimal feedback regime and for $\Delta\gg\gamma$, this ratio is
given by
\beq
\frac{\gamma\mu\Omega_{p}^{2}/4\Delta^{2}}{\kappa\mu}=
\frac{4\mu M}{\kappa}\frac{2AI_{\rm sat}}{\hbar\omega_p\gamma\mu}
\approx\chi\frac{2AI_{\rm sat}}{\hbar\omega_p\gamma\mu}.
\label{ratio}
\eeq
For the typical values stated above, \erf{ratio} is indeed
small ($\sim 10^{-1}$).

Another practical question is, how realistic is the
zero-time-delay assumption for the feedback? It can be shown using
the techniques of Ref.~\cite{Wis94} that this assumption is
justified providing the feedback delay time is much less than
$\kappa^{-1}$, the lifetime of the trap due to the output
coupling. If recent experiments \cite{pulses,beams} are a useful
guide, trap lifetimes are of order $10^{-2}$s \cite{WisVac02}.
Feedback much faster than this should not be a problem. In fact,
the time delay could be completely eliminated by feeding forward
rather than feeding back. Linewidth reduction can be achieved
equally well by controlling the phase of the atom field once it
has left the trap as by controlling it inside the trap (but, of
course, an integrated, rather than instantaneous, current would be
used for the control).

\section{Summary}

The coherence of an atom laser can be defined \cite{Wis97}
analogously to that of an optical laser: it should be
monochromatic with small intensity and phase fluctuations. We used
the normalized first-order coherence function
$g^{(1)}(t)=G^{(1)}(t)/\an{I}$ \cite{Glau63} as a measure of the
phase fluctuations. As $t$ increases, $|g^{(1)}(t)|$ decreases
from unity as the phase of the field becomes decorrelated from its
initial value. Its decay is characterized by the coherence time
$\tau_{\rm coh}=\frac{1}{2}\int_{0}^{\infty}|g^{(1)}(t)|dt$, or by
its reciprocal, the linewidth $\ell$. $G^{(1)}(t)$ also determines
the output power spectrum, where the peak spectral height is given
by $4\an{I}/\ell$ regardless of lineshape, while for a Lorentzian
the FWHM is also equal to $\ell$.

We examined the linewidth as a function of the dimensionless
atomic interaction strength, $\chi=4\mu C/\kappa$, where $\kappa$
is the output coupling rate and $C$ is the atomic self-energy.
There are four distinct regimes: the standard laser linewidth for
$\chi\ll1$, a quadratic dependence on $\chi$ for
$1\ll\chi\ll\sqrt{\mu}$, a linear regime for
$\sqrt{\mu}\ll\chi\ll\mu^{2}$, and finally another constant regime
when $\chi\gg\mu^{2}$. The second and third regimes have
Lorentzian and Gaussian output spectra respectively. The last
regime is a consequence of quantum revivals of $|g^{(1)}(t)|$,
which are a direct effect of strong atomic collisions
\cite{WriWalGar96}. This leads to a complicated structure in the
power spectrum with many peaks contained in a Gaussian-like
envelope.

An important condition for atom-laser coherence is that the phase
fluctuations be small in a particular sense. This is equivalent to
requiring Bose degeneracy in the output, i.e., that the linewidth
$\ell$ be much less than the the output flux $\an{I}=\kappa\mu$.
From the results presented here, this means that the laser output
is only coherent for interaction strengths satisfying
$\chi\alt\mu^{3/2}$, i.e., somewhere in the third linewidth
regime. Therefore, collisions will be a problem for atom-laser
coherence, especially for BECs formed by evaporative cooling where
collisions are the dominant mechanism. On the other hand, if the
atom-atom interactions are strong enough, the laser output will
exhibit the interesting feature of quantum revivals.

We also show, expanding upon Ref.~\cite{WisTho01}, that this
linewidth broadening can be significantly reduced by a QND
feedback scheme. Basically, by feeding back the results of a QND
measurement of the number fluctuations to control the condensate
energy, it is possible to compensate for the linewidth caused by
the frequency fluctuations. The very number-phase correlation
created by the collisions is utilized to cancel their effect. We
have shown that this linewidth reduction allows the output to
remain coherent for interaction strengths up to
$\chi\simeq\mu^{2}$ rather than $\mu^{3/2}$, which is an
improvement by a factor of $\sqrt{\mu}$. For the reasonable
parameters of $C\sim10^{-2}s^{-1}$ and $\mu\sim10^{6}$, this
improvement in linewidth is of the order of $10^{3}$ and in
principle could increase coherent values of $\chi=4\mu C/\kappa$
from $\sim10^{9}$ to $\sim10^{12}$.

\acknowledgments

H.M.W is deeply indebted to W. D. Phillips for the idea of
controlling atom-laser phase fluctuations using atom number
measurements.

\appendix

\section{Single complex exponential approximation for $g^{(1)}(t)$}

For the majority of the calculations in this paper we assume that
the first order coherence function can be given by \beq
g^{(1)}(t)=|g^{(1)}(t)|e^{i\bar{\omega}t}. \eeq That is, we assume
that the complex nature of $g^{(1)}(t)$ (which is due only to
collisions as shown in Sec.~III D) is described by a single
complex exponential for most regimes of the collisional
nonlinearity. When the collisions are strong enough to cause
revivals this approximation necessarily breaks down.

To determine the relative error caused by this approximation we
continue the analysis of evolution due to collisions presented in
Sec~III D, i.e., for $\dot{\rho}={\cal L}_{C}\rho$. Specifically,
we quantify the relative difference between \beq \tau_{\rm
exact}=\frac{1}{2}\int_{0}^{\infty}e^{-\mu(1-\cos2Ct)}dt \eeq and
\beq \tau_{\rm
approx}=\frac{1}{2}\int_{0}^{\infty}e^{-\mu[1-\exp(2iCt)]-i\bar{\omega}t}dt.
\eeq This is done by multiple Taylor expansions [first of
$\cos2Ct$ and $\exp(2iCt)$, and then expanding resultant
exponentials apart from $\exp(at^2)$] and using the equality \beq
\int_{0}^{\infty}t^{n}e^{-2\mu
C^{2}t^{2}}dt=\frac{\Gamma[(n+1)/2]}{2(2\mu C^{2})^{(n+1)/2}},
\eeq where $\Gamma[n]$ is the Gamma function.

Now, $\tau_{\rm approx}$ differs from $\tau_{\rm exact}$ by
both real and imaginary terms. The dominant
[for $t<(\mu C^2)^{-1/2}$] real term is
\beq
-\frac{(2\mu C-\bar{\omega})^{2}}{32}\sqrt{\frac{\pi}{2}}\ro{\frac{1}{\mu C^{2}}}^{3/2}
+... \label{rediff}
\eeq
while the first three imaginary terms are
\beq
\frac{(2\mu C-\bar{\omega})}{8\mu C^{2}}
-\sq{\frac{2}{3}\mu C^{3}+\frac{1}{12}(2\mu C-\bar{\omega})^{3}}\frac{1}{8(\mu C^{2})^{2}}
+... \label{imdiff}
\eeq
We can thus determine $\bar{\omega}$ by requiring that the
imaginary terms vanish. This is essentially what the iterative
procedure of Appendix B achieves.

The series of imaginary terms above leads to a first choice of
$\omega_{0}=2\mu C$, i.e., this sets the first term to zero and we
are left with terms of ${\cal O}[(\mu C)^{-1}]$. To cancel the
first two imaginary term we require $2\mu C-\bar{\omega}=2C/3$,
leaving terms of ${\cal O}[(\mu^2 C)^{-1}]$. Hence,
$\omega_{1}=2\mu C-2C/3$ will largely ensure that $\tau_{\rm
approx}$ is real (since $\mu\gg1$). The question now is, how much
different is $\tau_{\rm approx}$ from $\tau_{\rm exact}$ using
$\omega_{1}$?

The dominant term in the difference is simply found by
substituting $\omega_{1}$ into \erf{rediff}, giving \beq
\frac{1}{36\mu^{3/2}C}\sqrt{\frac{\pi}{2}}. \eeq The relative size
of this error term is obtained by comparing to $\tau_{\rm exact}$,
which using the Taylor expansion equals \beq \tau_{\rm
exact}\simeq\frac{1}{2}\int_{0}^{\infty}e^{-2\mu C^{2}t^{2}}dt
=\frac{1}{4\mu^{1/2}C}\sqrt{\frac{\pi}{2}}. \eeq This corresponds
to a relative error, given by $(\tau_{\rm approx}-\tau_{\rm
exact})/\tau_{\rm exact}$, of the order of $\mu^{-1}$. We have
also verified the size of this error numerically for the full
system dynamics by simulations of $|g^{(1)}(t)|$ and
$g^{(1)}(t)\exp(-i\omega_{1}t)$ (here $\omega_{1}$ is found in the
same way as shown in Appendix B). These results also confirmed
that the single complex exponential approximation breaks down when
$C\sim\kappa\pi\mu$ (or $\chi\sim4\pi\mu^2$), as expected.

\section{Iterative procedure for the coherence time $\tau_{\rm coh}$}

As stated in Sec.~III B, the coherence time can be evaluated
numerically using \erf{numtau}, where the first guess for
$\bar{\omega}$ is given by \erf{omega0}. Subsequent corrections to
$\bar{\omega}$ are found by applying the procedure: \bqa
&\tau_{k}&=~-{{\rm Tr}[a\dg({\cal L}-i\omega_{k})^{-1}a\rho_{\rm
ss}]}
/2\an{a\dg a}, \label{B1} \nn\\
&\omega_{k+1}&=~\omega_{k}-{\rm Im}(1/2\tau_{k}). \eqa At each
step we have \beq
2\tau_{k}=\int_{0}^{\infty}g^{(1)}(t)e^{-i\omega_{k}t}dt, \eeq
which is simply a reexpression of \erf{B1}. Then using the
assumption of Appendix A that $g^{(1)}(t)\approx
|g^{(1)}(t)|e^{i\bar{\omega}t}$, we have \beq
2\tau_{k}\approx\int_{0}^{\infty}|g^{(1)}(t)|e^{i(\bar{\omega}-\omega_{k})t}dt.
\eeq

The simplest form for $|g^{(1)}(t)|$ is a decaying exponential,
$\exp(-\gamma t)$, which gives rise to a Lorentzian power spectrum
as discussed at the end of Sec.~II. In this case \beq
2\tau_{k}\approx\int_{0}^{\infty}e^{-[\gamma-i(\bar{\omega}-\omega_{k})]t}dt
=\frac{1}{\gamma-i(\bar{\omega}-\omega_{k})}, \eeq and hence \beq
{\rm Im}(1/2\tau_{k})\approx\omega_{k}-\bar{\omega}. \label{B6}
\eeq The more complicated form of
$|g^{(1)}(t)|=\exp(-\gamma^{2}t^{2})$, which gives rise to a
Gaussian power spectrum, also obeys \erf{B6}, since in this case
\bqa
2\tau_{k}&\approx&\int_{0}^{\infty}e^{-[\gamma^{2}t^2-i(\bar{\omega}-\omega_{k})t]}dt \nn \\
&=&\frac{\sqrt{\pi}}{2\gamma}~e^{-(\bar{\omega}-\omega_{k})^{2}/4\gamma^{2}}
~{\rm erfc}\sq{\frac{-i(\bar{\omega}-\omega_{k})}{2\gamma}} \nn \\
&\simeq& -1/i(\bar{\omega}-\omega_{k}).
\eqa

Thus, if the coherence function is in the regime where the output
spectrum is Lorentzian or Gaussian, then
$\omega_{k+1}\approx\bar{\omega}$, and the first correction,
$\omega_{1}$, will be sufficient for an accurate calculation. On
the other hand, if the coherence function is in the regime of
revivals then the approximation $g^{(1)}(t)\approx
|g^{(1)}(t)|e^{i\bar{\omega}t}$ itself is no longer valid.
Therefore, the above numerical method will be wildly inaccurate
and we require another method for obtaining the coherence time and
linewidth. This is detailed in Sec.~III D.

\section{Q-function correspondence for saturated gain}

As stated in Sec. III C the master equation can be reexpressed as
a Fokker-Planck equation for a convenient probability
distribution. In this appendix we show that, for the Q function,
the operator correspondence for the gain term is given by
\erf{gainQ}. The individual superoperators in this expression have
the corresponding differential operators, \bqa
{\cal D}[a\dg]\rho &\to& -\dbd{n}n~Q(n,\varphi), \\
{\cal A}[a\dg]\rho &\to& \ro{n+\dbd{n}n} Q(n,\varphi). \eqa
Combining these leads to the following correspondence for the
saturated gain term: \beq {\cal D}[a\dg]{\cal
A}[a\dg]^{-1}\rho~\to-\dbd{n}n\ro{n+\dbd{n}n}^{-1}Q(n,\varphi),
\eeq which can be expanded in two different ways to give \beq
\cu{\ro{1+\frac{1}{n}\dbd{n}n}^{-1}-1}Q(n,\varphi), \eeq or \beq
-\dbd{n}\ro{1+\frac{1}{n}\dbd{n}n}^{-1}Q(n,\varphi). \eeq Equating
these expressions leads to \beq
\ro{1+\frac{1}{n}\dbd{n}n}^{-1}=\ro{1+\dbd{n}}^{-1}, \eeq and
hence \beq {\cal D}[a\dg]{\cal A}[a\dg]^{-1}\rho
~\to~\cu{\ro{1+\dbd{n}}^{-1}-1}Q(n,\varphi). \eeq By then applying
the Taylor expansion, $(1+x)^{-1}=\sum_{k=0}^{\infty}(-x)^{k}$, we
obtain the identity \beq {\cal D}[a\dg]{\cal A}[a\dg]^{-1}\rho
~\to~\sum_{k=1}^{\infty}\ro{-\dbd{n}}^{k}Q(n,\varphi). \eeq


\begin{thebibliography}{}
\bibitem{fn-continuous}
Of course, some optical lasers have a pulsed rather than
continuous output. However, as argued in Ref.~\cite{Wis97}, in
order for the term ``atom laser'' to mean something more than
merely an atomic Bose-Einstein condensate released from its trap,
it is useful to impose the condition that a true atom laser has a
continuous output.
\bibitem{fn-pspec}
Based on an input-output formalism for atoms, the time dependence
of the output spectrum has also been calculated \cite{inout97},
where the power spectrum defined in \erf{PS} is simply the
steady-state limit.
\bibitem{fn-compmem}
Larger values of $\mu$ have prohibitive computer memory
requirements.
\bibitem{fn-antimean}
For simplicity, we will assume that the distortion of the beam
front, and the mean phase shift, are removed by a suitable
``anti-mean-condensate'' lens.
\bibitem{fn-depletion}
The depletion rate scales as $M$, so this will not be negligible
compared to the output loss rate when $M\agt\kappa$. For the
optimal regime, this would occur when $C\agt\kappa$, that is
$\chi\agt\mu$. As argued in \cite{WisVac02}, this is not the
expected regime of operation for an atom laser. Nevertheless, it
means that the higher regions of the feedback curve in Fig.~2
should be taken with a grain of salt.
\end{thebibliography}
\end{document}